\documentclass[aps,prx,superscriptaddress,twocolumn,longbibliography]{revtex4-1}

\usepackage{amsfonts}
\usepackage{subfigure}
\usepackage{amsmath}
\usepackage{txfonts}
\usepackage{amssymb}
\usepackage{amsbsy} 
\usepackage{epsfig}
\usepackage{graphicx}

\usepackage{float} 

\def\be{\begin{equation}} \def\ee{\end{equation}}
\def\bea{\begin{eqnarray}} \def\eea{\end{eqnarray}}

\def\nn{\nonumber}

\def\bk{{\bf k}}

\def\be{{\bf e}}

\def\bN{{\bf N}}
\def\bn{{\bf n}}

\begin{document}

\title{Nodal-knot semimetals}

\author{Ren Bi}
\affiliation{ Institute for
Advanced Study, Tsinghua University, Beijing, 100084,  China}

\author{Zhongbo Yan}
\affiliation{ Institute for
Advanced Study, Tsinghua University, Beijing, 100084,  China}

\author{Ling Lu}
\affiliation{ Institute of Physics, Chinese Academy of Sciences/Beijing National Laboratory for Condensed Matter Physics, Beijing 100190, China }

\author{Zhong Wang}
\altaffiliation{  wangzhongemail@tsinghua.edu.cn} \affiliation{ Institute for
Advanced Study, Tsinghua University, Beijing, 100084,  China}
\affiliation{Collaborative Innovation Center of Quantum Matter, Beijing, 100871, China }


\begin{abstract}

Topological nodal-line semimetals are characterized by one-dimensional lines of band crossing in the Brillouin zone. In contrast to nodal points, nodal lines can be in topologically nontrivial configurations. In this paper, we study the simplest topologically nontrivial forms of nodal line, namely, a single nodal line taking the shape of a knot in the Brillouin zone. We introduce a generic construction for various ``nodal-knot semimetals'', which yields the simplest trefoil nodal knot and other more complicated nodal knots in the Brillouin zone. The knotted-unknotted transitions by nodal-line reconnections are also studied. Our work brings the knot theory to the subject of topological semimetals.

\end{abstract}

\maketitle

\section{Introduction}

Topological states have been under intense investigations in the last decade\cite{hasan2010,qi2011,Bansil2016, bernevig2013topological,shen2013topological}.
Topological semimetals\cite{Chiu2015RMP} are characterized by topologically protected nodal points or nodal lines in the Brillouin zone, where the valance band and conduction band meet each other.  The most extensively studied nodal-point semimetals are Weyl semimetals\cite{wan2011,murakami2008,burkov2011,yang2011,volovik2003,
weng2015,Huang2015TaAs,Xu2015weyl,lv2015,Huang2015,Xu2015NbAs,
YangLexian,Shekhar,lu2015,soluyanov2015type} and Dirac semimetals\cite{neupane2014,xu2015observation,liu2014discovery,Borisenko2014, young2012dirac,wang2012dirac,wang2013three,
Chen2015Magnetoinfrared,liu2016zeeman}. More recently, nodal-line semimetals have attracted considerable attention. Like the Dirac points, nodal lines\cite{Burkov2011nodal,
Phillips2014tunable,Carter2012,Chiu2014,
Zeng2015nodal,chen2015topological,Weng2015nodal,Bian2015nodal,Yu2015,Kim2015,
Rhim2015Landau,Chen2015spin,
Fang2015nodal,Mullen2015,Bian2015TlTaSe,yu2017topological} are protected by both the band topology and symmetries. Nodal-line semimetals are versatile platforms for topological materials.
By breaking certain symmetry, nodal lines can be partially or fully gapped, giving way to Dirac semimetals, Weyl semimetals, or topological insulators. These phenomena can be triggered by the spin-orbit coupling\cite{Yu2015,Kim2015} or external driving\cite{yan2016tunable,Chan2016type, Narayan2016nodal,Zhang2016floquet,Taguchi2016nodal}. There are quite a few material candidates of nodal lines: Cu$_3$NPd and Cu$_3$NZn\cite{Yu2015,Kim2015}, calcium phosphide\cite{Chan2015Ca3P2,Xie2015ring}, carbon networks\cite{Weng2015nodal}, ${\mathrm{CaP}}_{3}$\cite{Xu2017nodal}, alkali earth materials\cite{Li2016alkai,Hirayama2017nodal}, to name a few. Experimental studies of nodal lines are also in rapid progress\cite{Bian2015nodal,schoop2015dirac,hu2016ZrSiTe, Singha2016,Neupane2016,Wang2016evidence,Chen2017Dirac}.

\begin{figure}
\subfigure{\includegraphics[width=4cm, height=4cm]{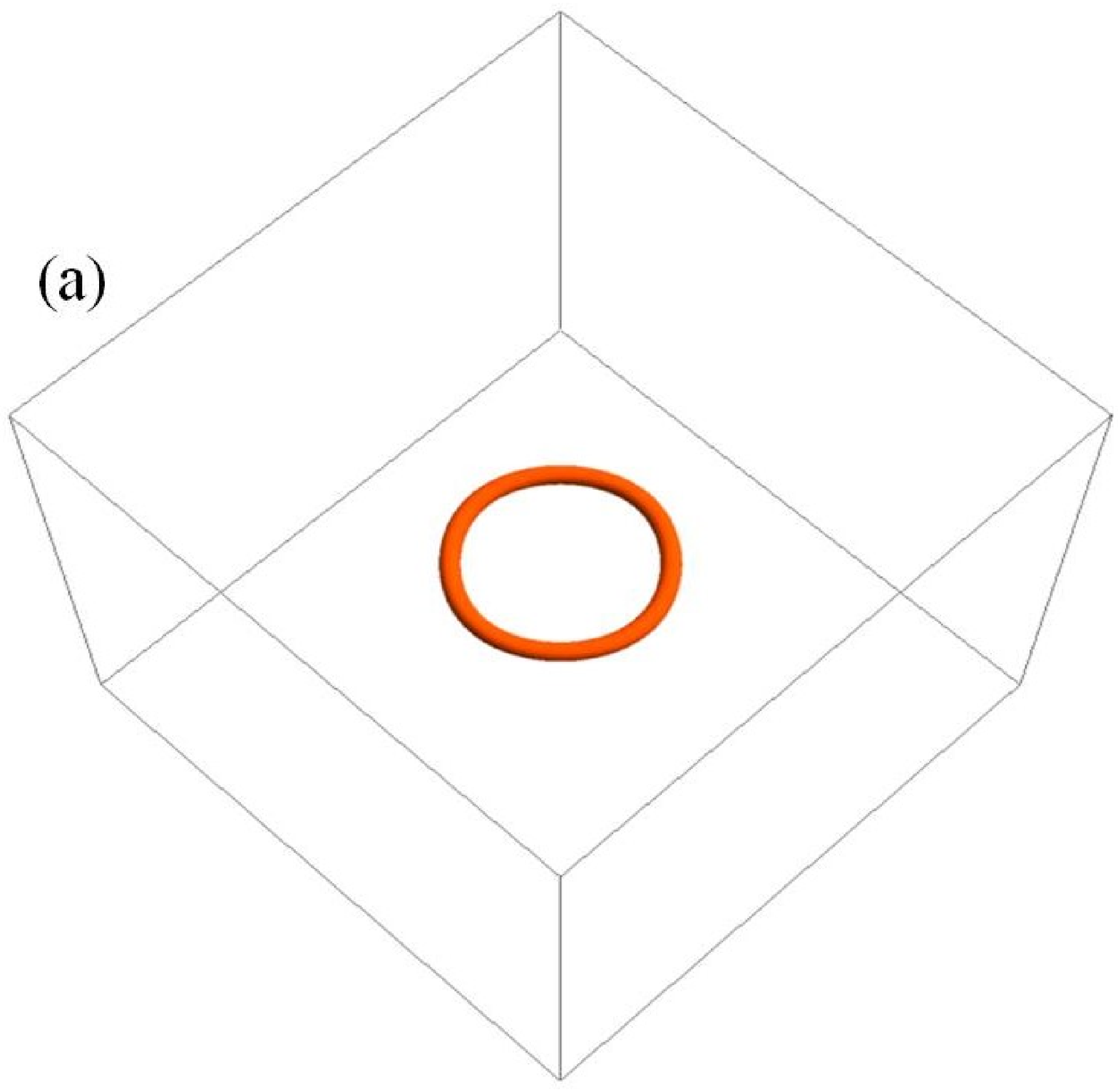}}
\subfigure{\includegraphics[width=4cm, height=4cm]{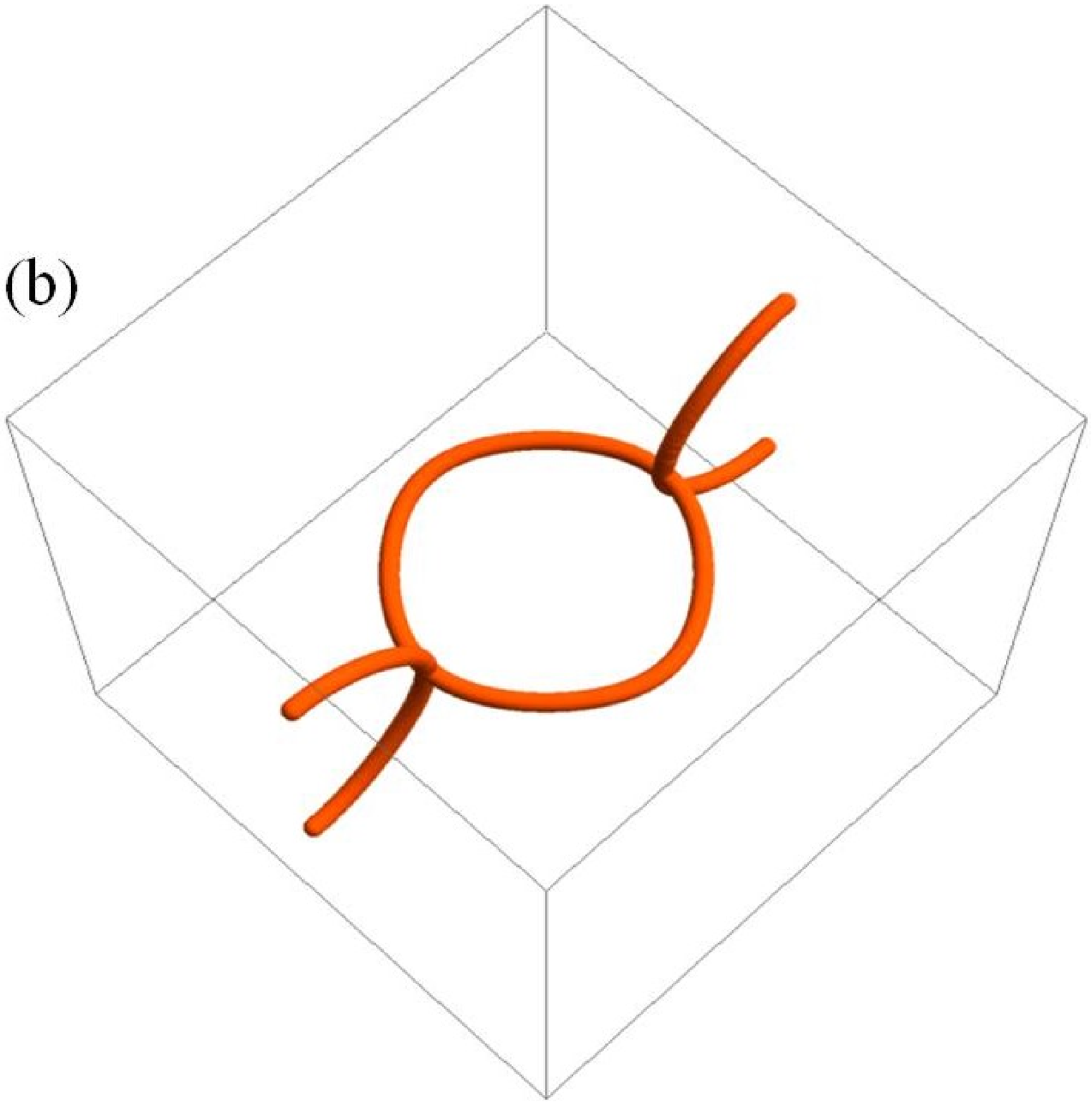}}
\subfigure{\includegraphics[width=4cm, height=4cm]{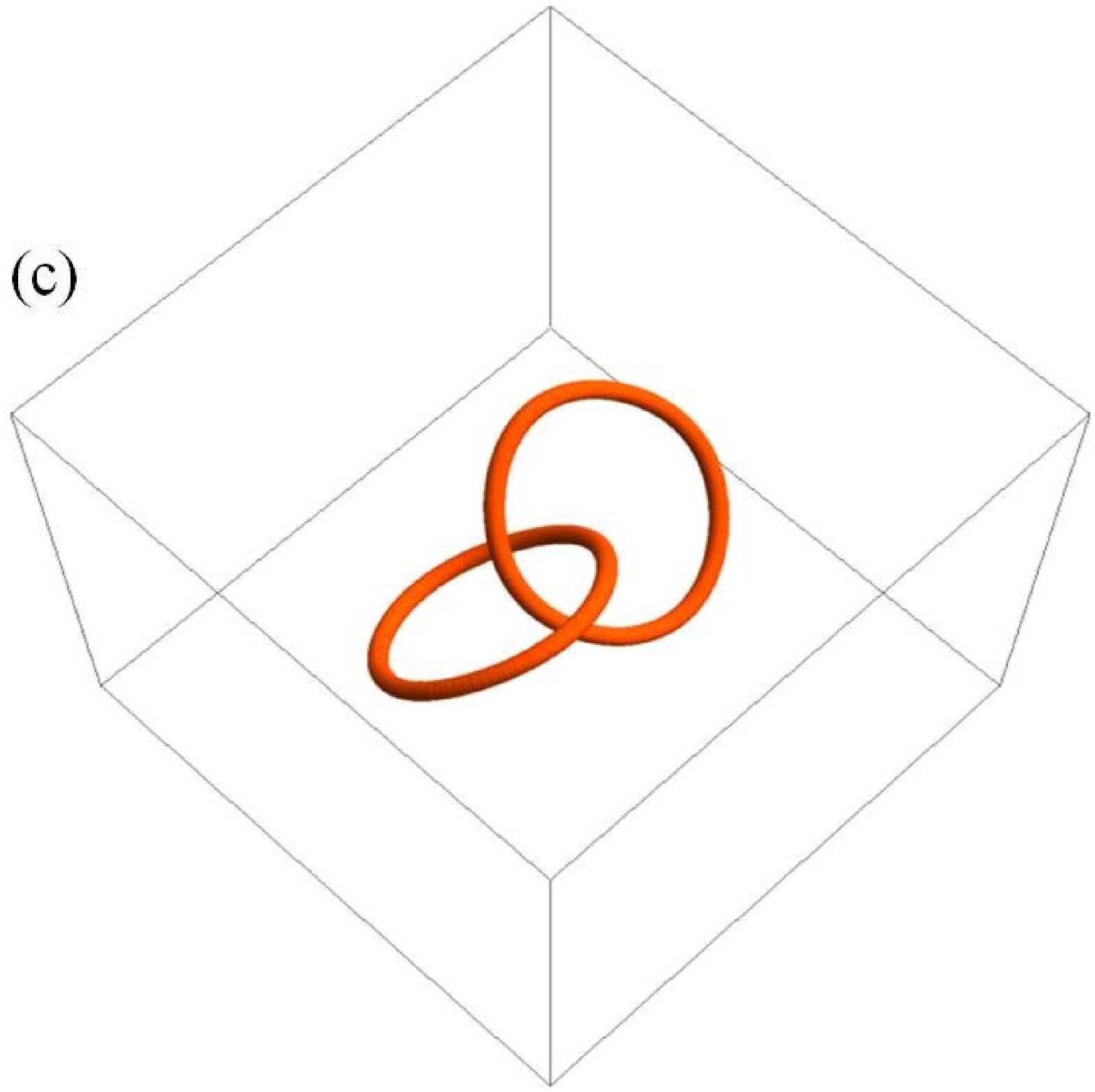}}
\subfigure{\includegraphics[width=4cm, height=4cm]{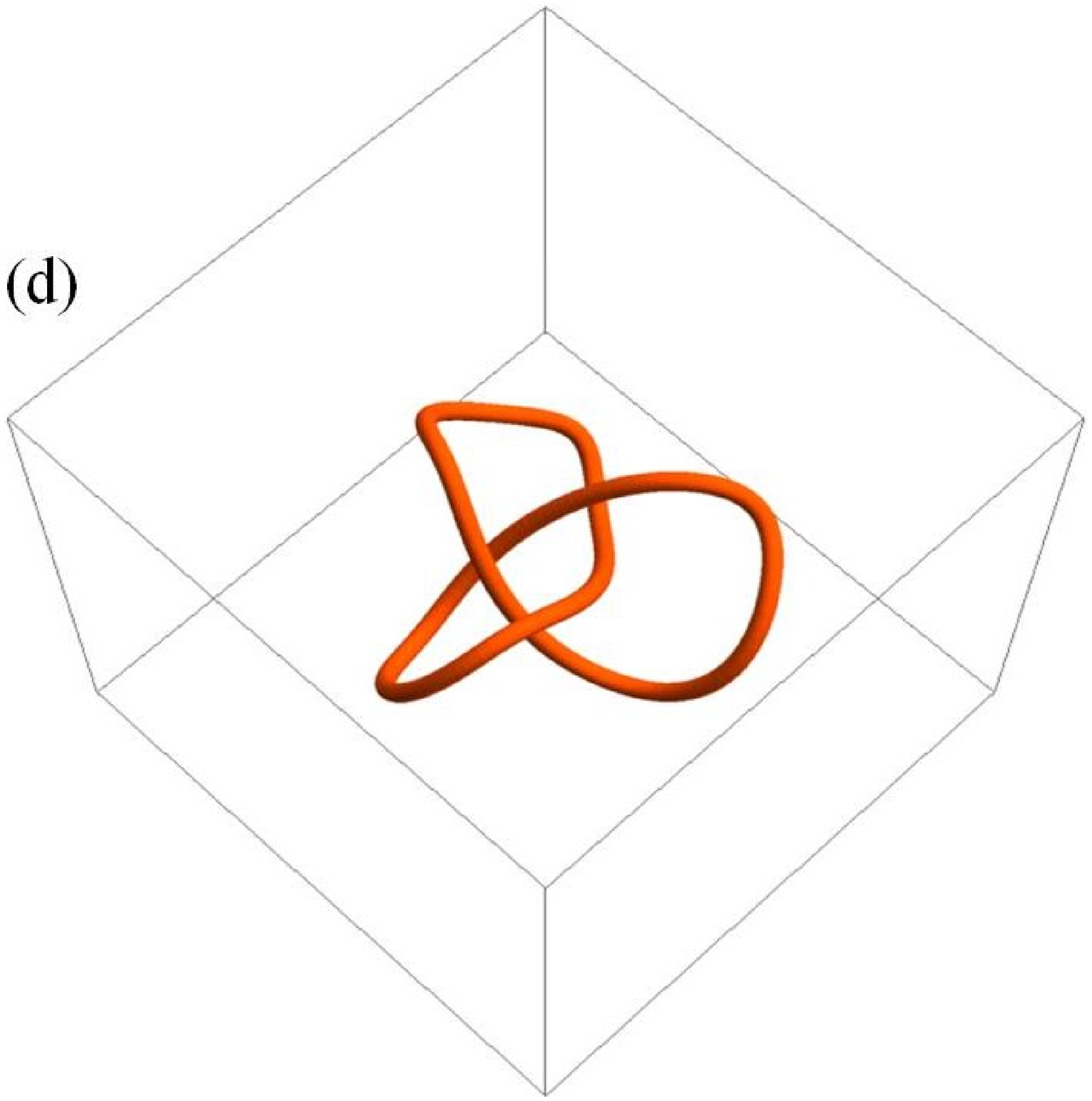}}
\caption{ Schematic illustration of four types of nodal line: (a) ordinary nodal ring, (b) nodal chain, (c) nodal link, and (d) nodal knot. A nodal knot is a single nodal line entangled with itself.  }  \label{sketch}
\end{figure}

Nodal points have little internal structure, for instance, the only topological characterization of a Weyl point is its chirality ($\pm 1$).  In contrast, nodal lines can have much richer topologically distinct possibilities. They can touch each other and form nodal chains stretching across the Brillouin zone\cite{Bzdusek2016,Yu2017chain} [Fig.\ref{sketch}(b)]. Another intriguing possibility is the nodal link\cite{hopflink,nodal-link,weyl-link}, namely, two nodal lines topologically linked with each other [Fig.\ref{sketch}(c)]. Links are also proposed in superconductors\cite{Sun2017helix}.

Nodal links are not the simplest topologically nontrivial shapes of nodal lines. The simplest shape contains only one nodal line entangled with itself, i.e., a knot. Such a nontrivial nodal line is dubbed a ``nodal knot'' in this paper, to distinguish it from the usual real-space knots. The simplest case is a trefoil nodal knot, shown in  Fig.\ref{sketch}(d). A crucial question is whether such a conception can exist in principle (i.e., can be explicitly constructed as models). The recently proposed method of constructing nodal links based on Hopf mappings cannot be applied to yield a nodal knot, because it necessarily  produces multiple nodal lines. It is thus unclear whether a single-line nodal knot is realizable in materials. Here, we introduce a method based on functions of several complex variables, which neatly gives various single-line nodal knots, including the trefoil knot as the simplest case. The topological transitions from the knotted configurations to trivial ones are also studied. Our work brings the extensively investigated knot theory\cite{kauffman2001knots} to the subject of topological semimetals.

\section{Continuum models of nodal-knot semimetals}

Nodal lines come from the crossing of two adjacent bands, thus we focus on two-band models below. Any two-band model can be written as \bea H(\bk)= a_1(\bk) \sigma_x +  a_2(\bk)\sigma_y + a_3(\bk)\sigma_z + a_0(\bk){\bf 1}, \eea
where $\sigma_i$'s are the Pauli matrices, and the trivial $a_0(\bk)$ term will be discarded below. Nodal lines are protected by the cooperation of band topology and certain symmetries\cite{horava2005,Zhao2013classification,Zhao2016Unified}. In this work, we consider the combination\cite{Fang2015nodal,Zhao2017PT} of spatial-inversion symmetry $\mathcal{P}$ and time-reversal symmetry $\mathcal{T}$, which ensures that $H^*(\bk)=H(\bk)$ up to a basis choice, thus we have $a_2(\bk)=0$. This $\mathcal{PT}$ symmetry is relevant to many material candidates of nodal lines (e.g., Cu$_3$PdN\cite{Kim2015,Yu2015}, Cu$_3$TeO$_6$\cite{li2017dirac}). Given the symmetry, the Hamiltonian becomes \bea H(\bk)=  a_1(\bk)\sigma_x + a_3(\bk)\sigma_z, \label{general} \eea whose energies are $E_\pm(\bk)=\pm\sqrt{a_1^2(\bk)+a_3^2(\bk)}$, and the nodal lines can be solved from $a_1(\bk)=a_3(\bk)=0$. The most common choices of $a_1$ and $a_3$, say $a_1=\cos k_x+\cos k_y +\cos k_z -m_0$, $a_3=\sin k_z$, yield ordinary nodal lines resembling  Fig.\ref{sketch}(a). Taking advantage of the Hopf mappings\cite{wilczek1983,nakahara2003,moore2008topological, deng2013hopf,deng2016probe,deng2015systematic, kennedy2016,wang2016measuring,liu2016symmetry}, nodal links such as the one shown in Fig.\ref{sketch}(c) can be constructed\cite{nodal-link}. This method is sufficiently general to generate nodal links with any integer linking numbers (including the simplest Hopf link), however, it is unable to generate a nodal knot, which contains only one nodal line. This limitation is intrinsic in its construction\cite{nodal-link}.

In this paper, we introduce a generic approach for constructing nodal knots, which is based on functions of several complex variables. Before writing down the explicit models, we start from the geometrical preparations.  Let us consider two complex variables $z$ and $w$, with the constraint $|z|^2+|w|^2=1$, which defines a 3-sphere.
This is more transparent if we write $z=n_1+in_2$ and $w=n_3+in_4$, then $|z|^2+|w|^2=1$ becomes $n_1^2+n_2^2+n_3^2+n_4^2=1$, which is apparently a 3-sphere. For reasons to become clear shortly, let us consider the surface $|z|^p=|w|^q$ (where $p,q$ are positive integers) in the 3-sphere. This surface is topologically a 2-torus. To see this fact, we notice that the two equations $|z|^2+|w|^2=1$ and $|z|^p=|w|^q$ completely fix the values of $|z|$ and $|w|$, thus the surface can be parameterized by the phases $\theta_z$ and $\theta_w$, which is defined in $z=|z|\exp(i\theta_z)$ and $w=|w|\exp(i\theta_w)$, respectively. Thus the surface is exactly a 2-torus, with $\theta_z$ and $\theta_w$ parameterizing the toroidal and poloidal direction, respectively.

Now we impose a constraint \bea f(z,w)\equiv z^p+w^q=0. \label{key} \eea   The solutions $(z,w)$ of this constraint must be on the torus $|z|^p=|w|^q$ discussed above, furthermore, the phases have to satisfy $p\theta_z-q\theta_w=\pi$ (mod $2\pi$). As a mathematical fact of torus geometry, when $p$ and $q$ are relatively prime, the equation defines only one line on the torus. Otherwise, we have multiple lines. For instance, when $(p,q)=(3,2)$, we have a single line passing the point $(\theta_z,\theta_w)=(\pi/3,0)$; when $(p,q)=(2,4)$, we have two disconnected lines, one of which passes $(\theta_z,\theta_w)=(\pi/2,0)$ and the other passes $(\theta_z,\theta_w)=(\pi/2,\pi/2)$. Most notably, when both $p$ and $q$ are nonzero, the line(s) winds around the torus in both the toroidal and poloidal directions, forming a knot when $(p,q)$ are relatively prime, or links otherwise.

\begin{figure}
\subfigure{\includegraphics[width=6cm, height=5.5cm]{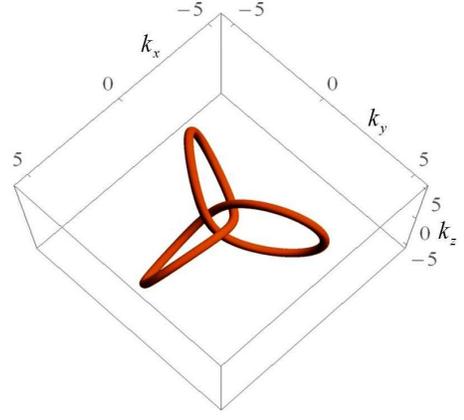}}
\caption{  The trefoil nodal knot of the continuum model in Eq.(\ref{continuum32}). The parameter $m$ in Eq.(\ref{continuum32}) is $m=0.5$.  }  \label{continuum} \end{figure}

With these geometrical preparations, we are now ready to construct continuum models of nodal-knot semimetals. For continuum models, the momentum variables $\bk=(k_x,k_y,k_z)$ extend to infinity. In the above construction of knot, the standard 3-sphere $n_1^2+n_2^2+n_3^2+n_4^2=1$ is considered. To make use of this construction, we can compactify the $\bk$-space to a 3-sphere by adding an ``infinity point'' (This is a standard procedure in topology, which is essentially the converse of stereographic projection).  We may establish a one-to-one correspondence between the compactified $\bk$-space and the standard 3-sphere. There are infinitely many ways to do this; for instance, we can take
\bea N_1= k_x,\, N_2= k_y,\, N_3=  k_z, \, N_4&=& m-k^2/2, \label{N} \eea
with $k^2=k_x^2+k_y^2+k_z^2$ and $m>0$, and define $n_i=N_i/N$ with $N=\sqrt{N_1^2+N_2^2+N_3^2+N_4^2}$. Now $\bn(\bk)=(n_1,n_2,n_3,n_4)$ maps the compactified $\bk$-space to the standard 3-sphere. It maps the origin $\bk=(0,0,0)$ to the north pole $\bn=(0,0,0,1)$, and the $\bk$-infinity to the south pole $\bn=(0,0,0,-1)$. The winding number of the mapping is known as \bea W= \frac{1}{2\pi^2}\int dk_x dk_y dk_z\,\epsilon^{abcd}n_a\partial_{k_x}n_b \partial_{k_y} n_c \partial_{k_z}n_d, \eea which is found to be $-1$ here. This is intuitively clear since the 3-sphere is covered only once. In fact, any other mapping with a nonzero winding number is applicable for our construction.

\begin{figure*}
\subfigure{\includegraphics[width=5.35cm, height=4.6cm]{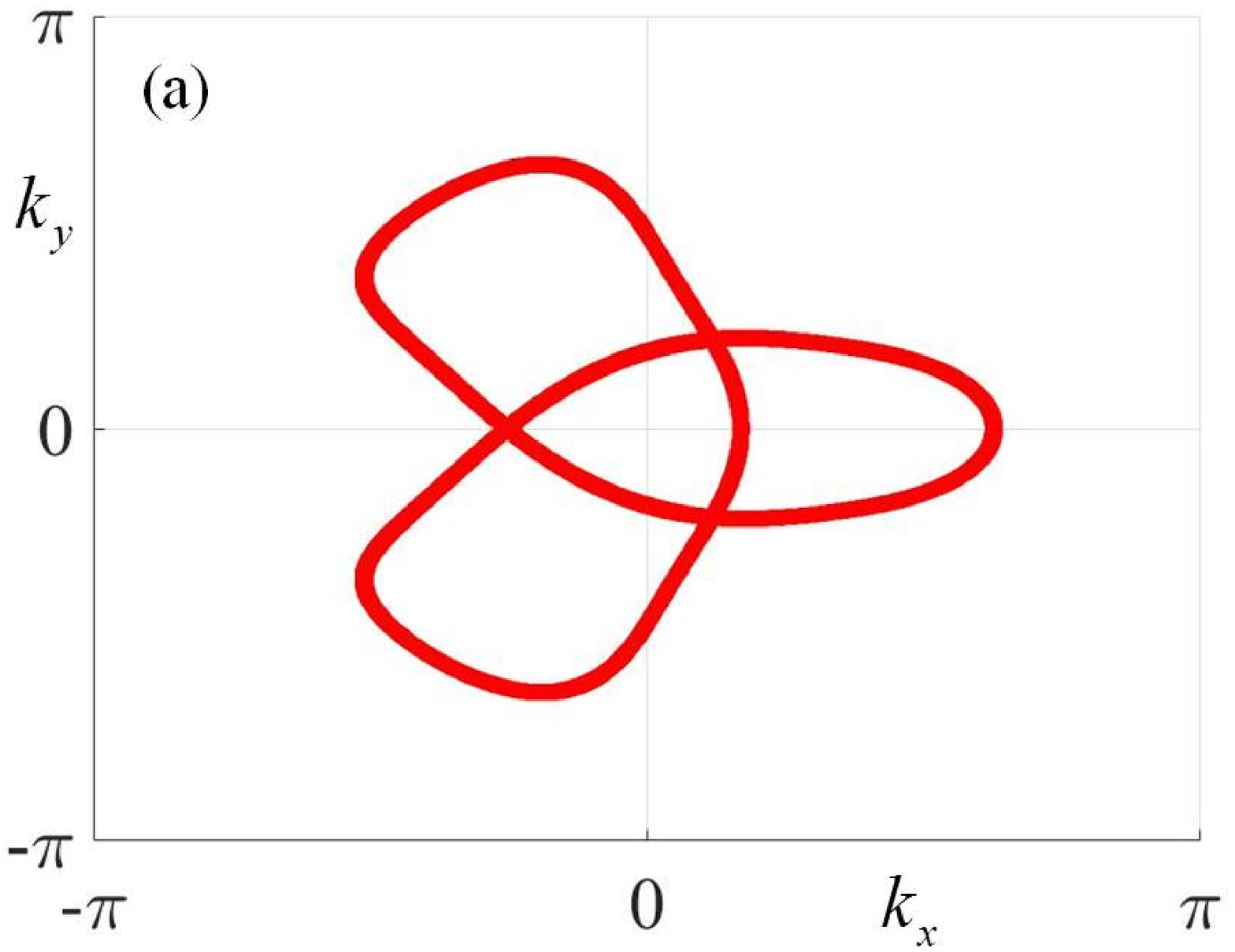}}
\subfigure{\includegraphics[width=5.35cm, height=4.6cm]{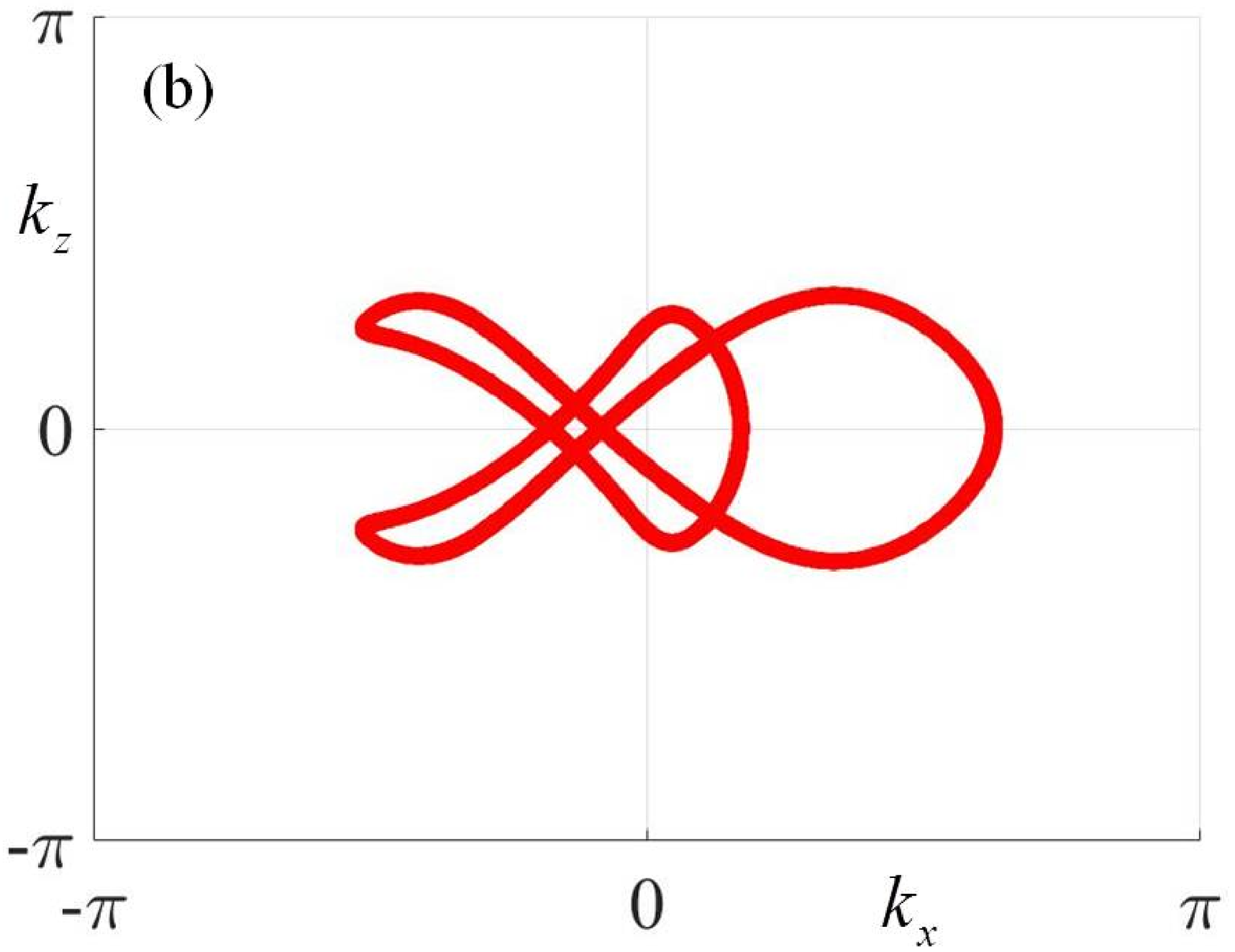}}
\subfigure{\includegraphics[width=5.35cm, height=4.6cm]{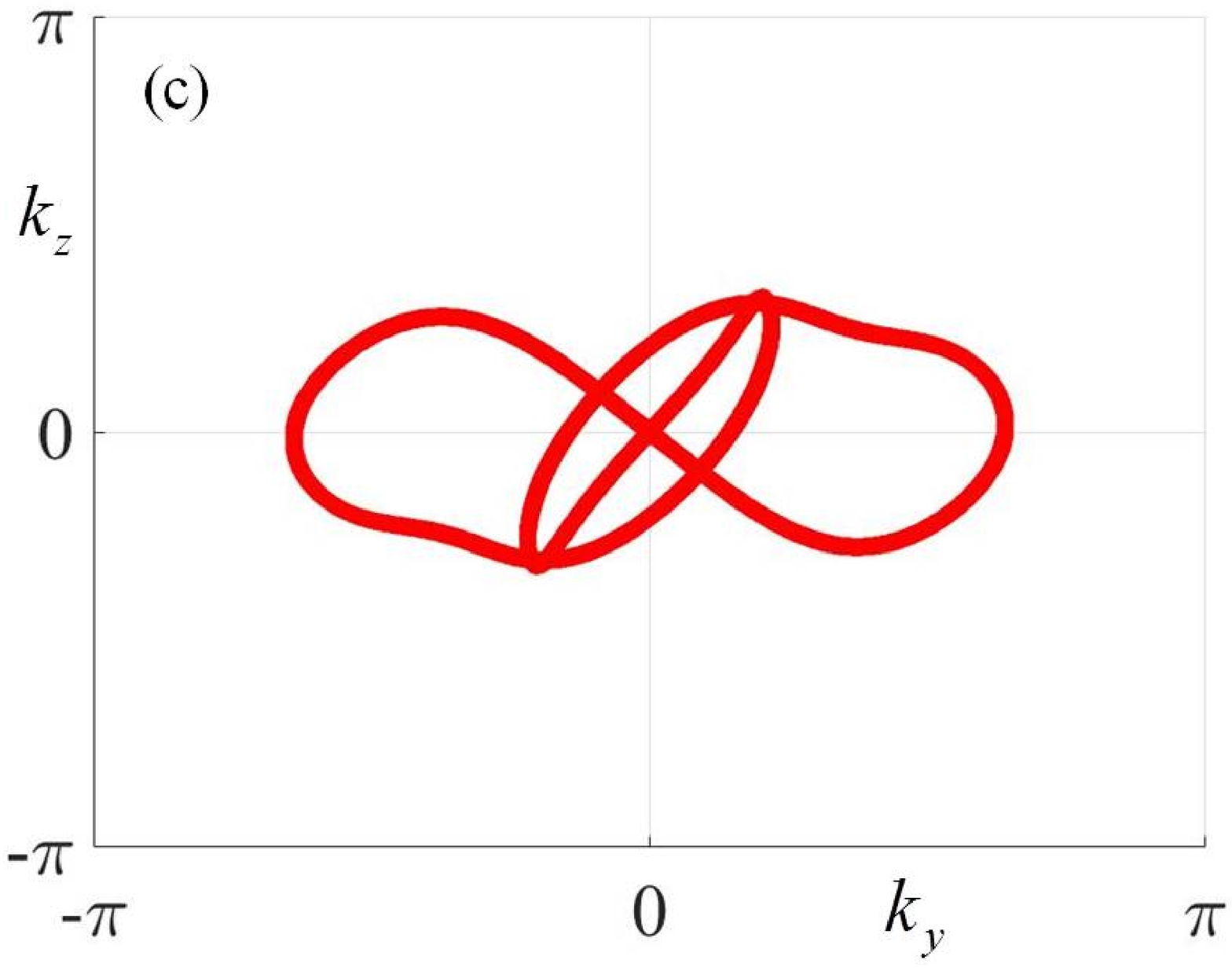}}
\subfigure{\includegraphics[width=5.35cm, height=4.6cm]{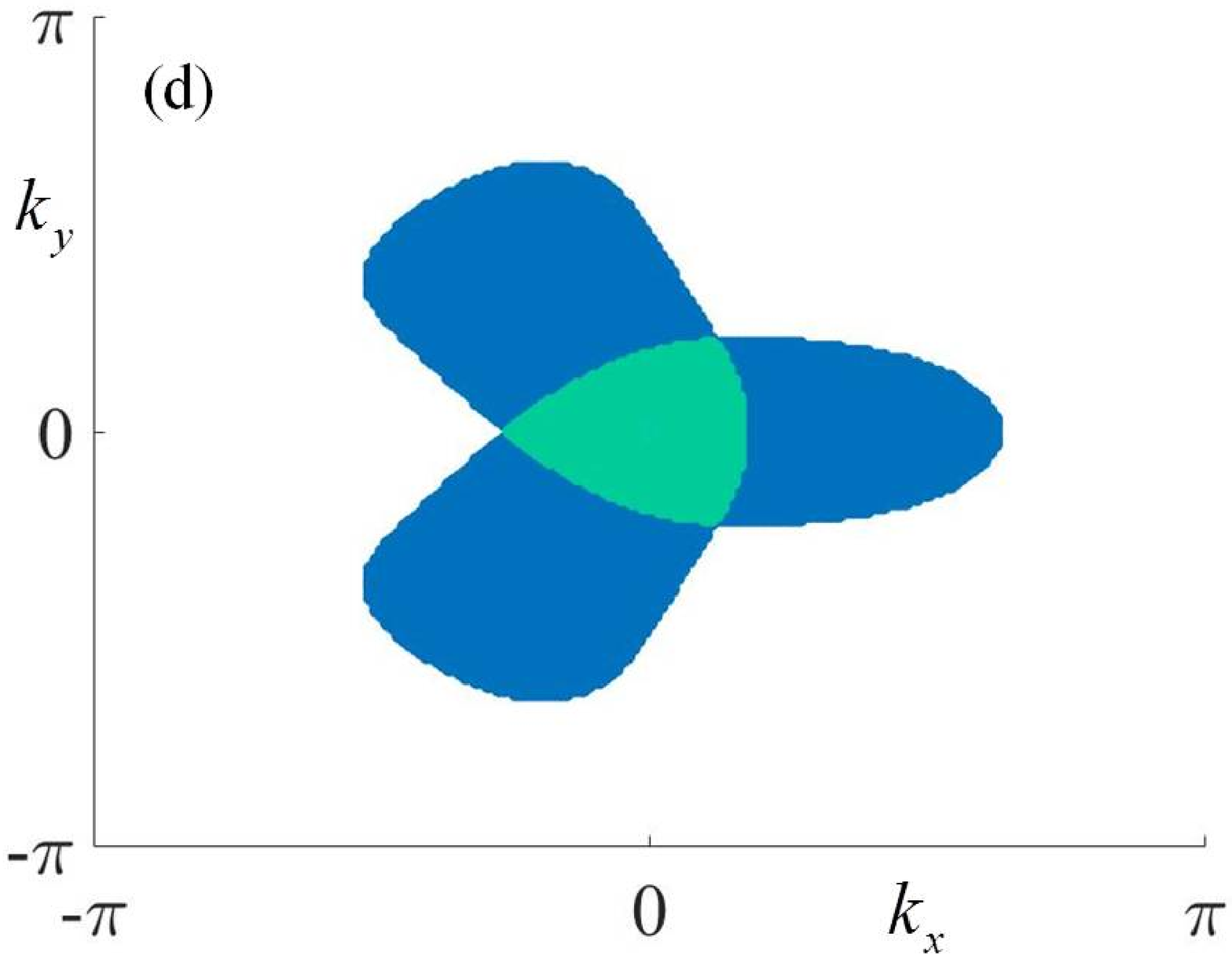}}
\subfigure{\includegraphics[width=5.35cm, height=4.6cm]{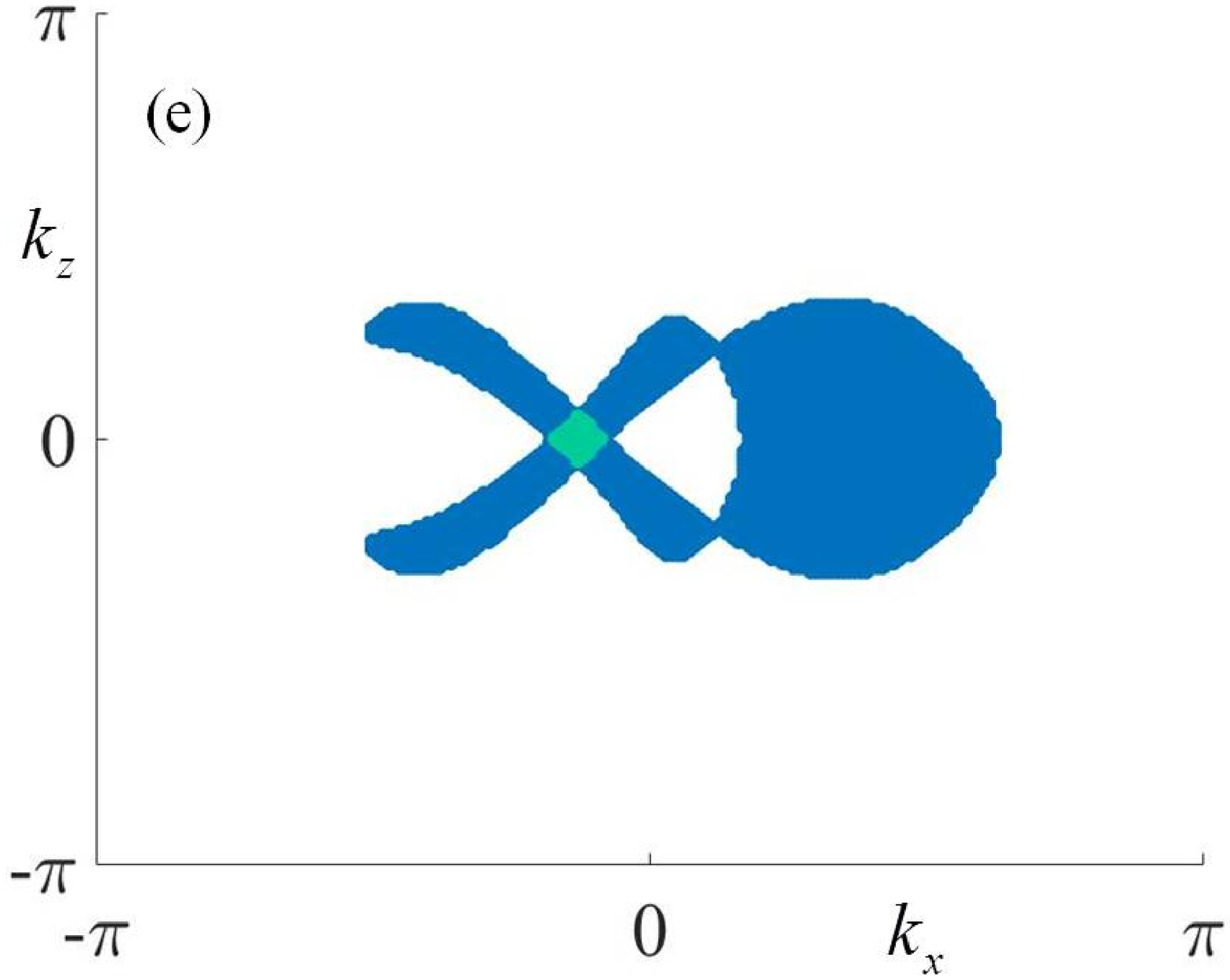}}
\subfigure{\includegraphics[width=5.35cm, height=4.6cm]{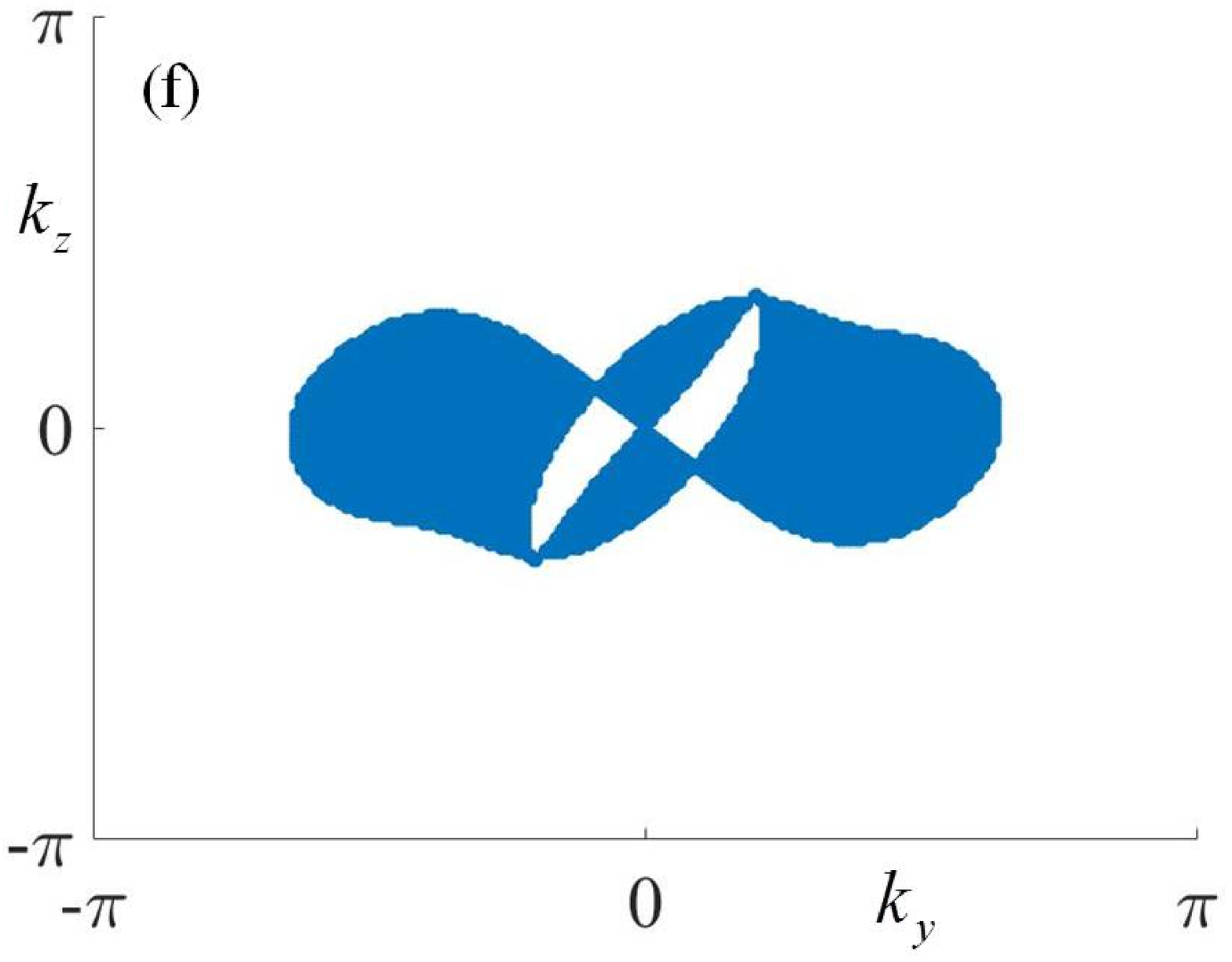}}
\caption{ (a)(b)(c): Nodal knots projected to the surface Brillouin zone, and (d)(e)(f): the regions of surface states, for the three spatial directions. In the blue regions in (d),(e), and (f), there is one surface band; in the green regions, there are two. The Bloch Hamiltonian is Eq.(\ref{general}) with $a_1,a_3$ given by Eq.(\ref{latticeH}); $m_0=2.5$. }  \label{latticesurface}
\end{figure*}

Now that $z=n_1+in_2$ and $w=n_3+in_4$ have become functions of $\bk$, we can take the coefficients $a_1$ and $a_3$ in Eq.(\ref{general}) as functions of $z$ and $w$. A natural choice is the real part and imaginary part of $f(z,w)$, respectively: \bea a_1(\bk) ={\rm Re} f(z,w), \quad a_3(\bk) = {\rm Im} f(z,w). \label{ansatz} \eea Eq.(\ref{key}) and Eq.(\ref{ansatz}) are among the key equations of this paper. With this ansatz, the nodal line equation $a_1(\bk)=a_3(\bk)=0$ is simply $f(z,w)=0$, i.e., Eq.(\ref{key}), which gives rise to a knot when $(p,q)$ are relatively prime, as explained above. This is the motivation of the ansatz. To be simpler, we can take \bea  z=N_1+iN_2,\quad w=N_3+iN_4  \label{convention} \eea in Eq.(\ref{ansatz}), which is topologically equivalent to taking $z=n_1+in_2,w=n_3+in_4$, because $\bn(\bk)$ and $\bN(\bk)$ differ only by a numerical factor $N(\bk)$. Hereafter we take the convention Eq.(\ref{convention}).

In the case $(p,q)=(3,2)$, we have \bea f(z,w)=(N_1+iN_2)^3+(N_3+iN_4)^2, \eea and the ansatz in Eq.(\ref{ansatz}) leads to
\bea a_1(\bk)&=& N_1^3 -3N_1 N_2^2 +N_3^2 -N_4^2,\nn\\ a_3(\bk)&=& 3N_1^2 N_2 -N_2^3 +2N_3 N_4, \eea
thus the Bloch Hamiltonian reads
\begin{eqnarray}
H(\bk)&=&[k_{x}^{3}-3k_{x}k_{y}^{2} + k_{z}^{2}-(m-k^{2}/2)^{2}]\sigma_{x}\nonumber\\
&&+[3k_{x}^{2}k_{y}-k_{y}^{3} +2k_{z}(m-k^{2}/2)]\sigma_{z}, \label{continuum32}
\end{eqnarray} where $k^2=\sum_{i=x,y,z} k_i^2$.  The nodal line of this model is shown in Fig.\ref{continuum}, which is apparently a trefoil nodal knot. Many other nodal knots can be obtained in this way by taking other $(p,q)$ or $\bN(\bk)$ function.

\section{Lattice models of nodal-knot semimetals and knotted-unknotted  topological transitions}

\begin{figure}[!h]
\subfigure{\includegraphics[width=4cm, height=3.5cm]{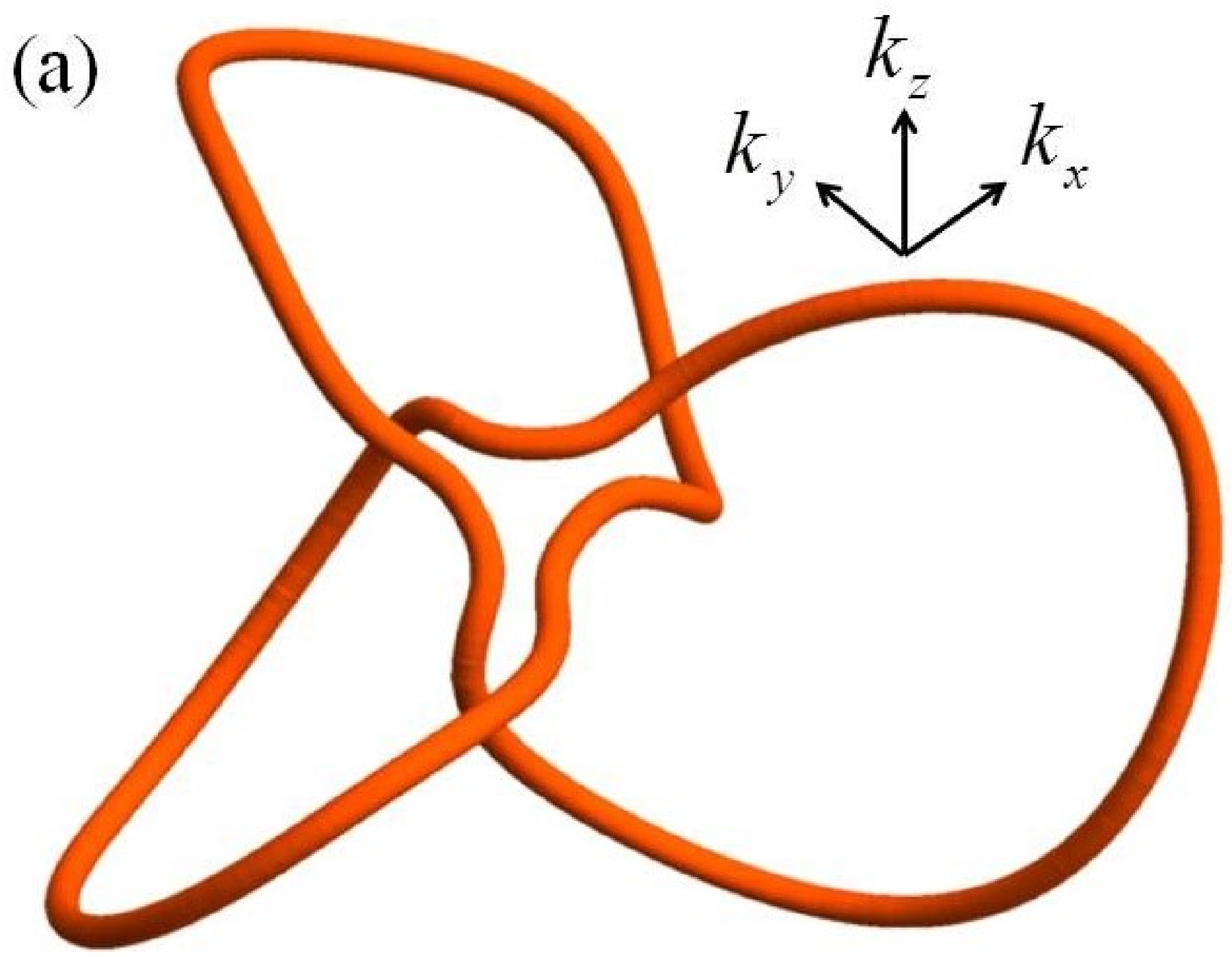}}
\subfigure{\includegraphics[width=4cm, height=3.5cm]{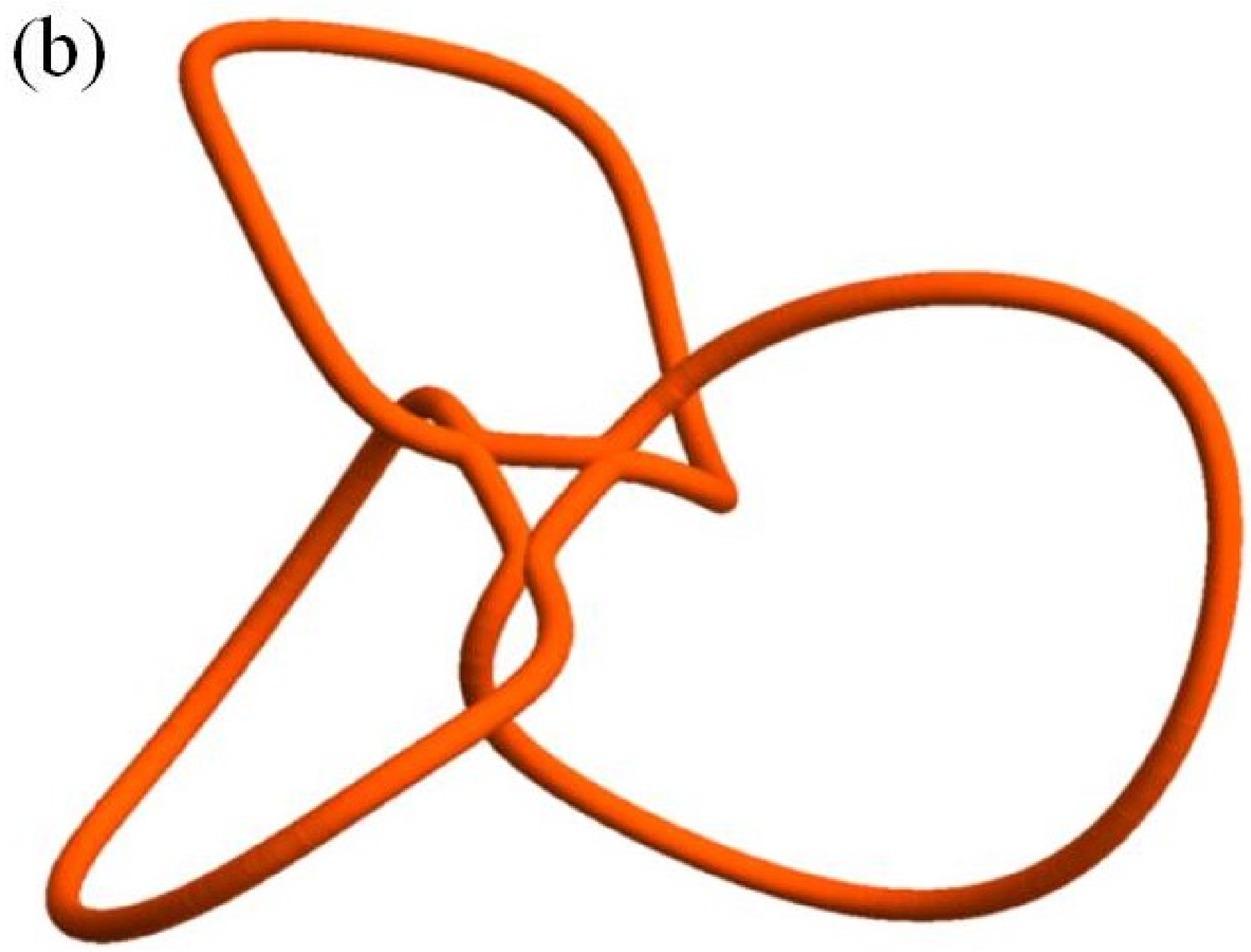}}
\subfigure{\includegraphics[width=4cm, height=3.5cm]{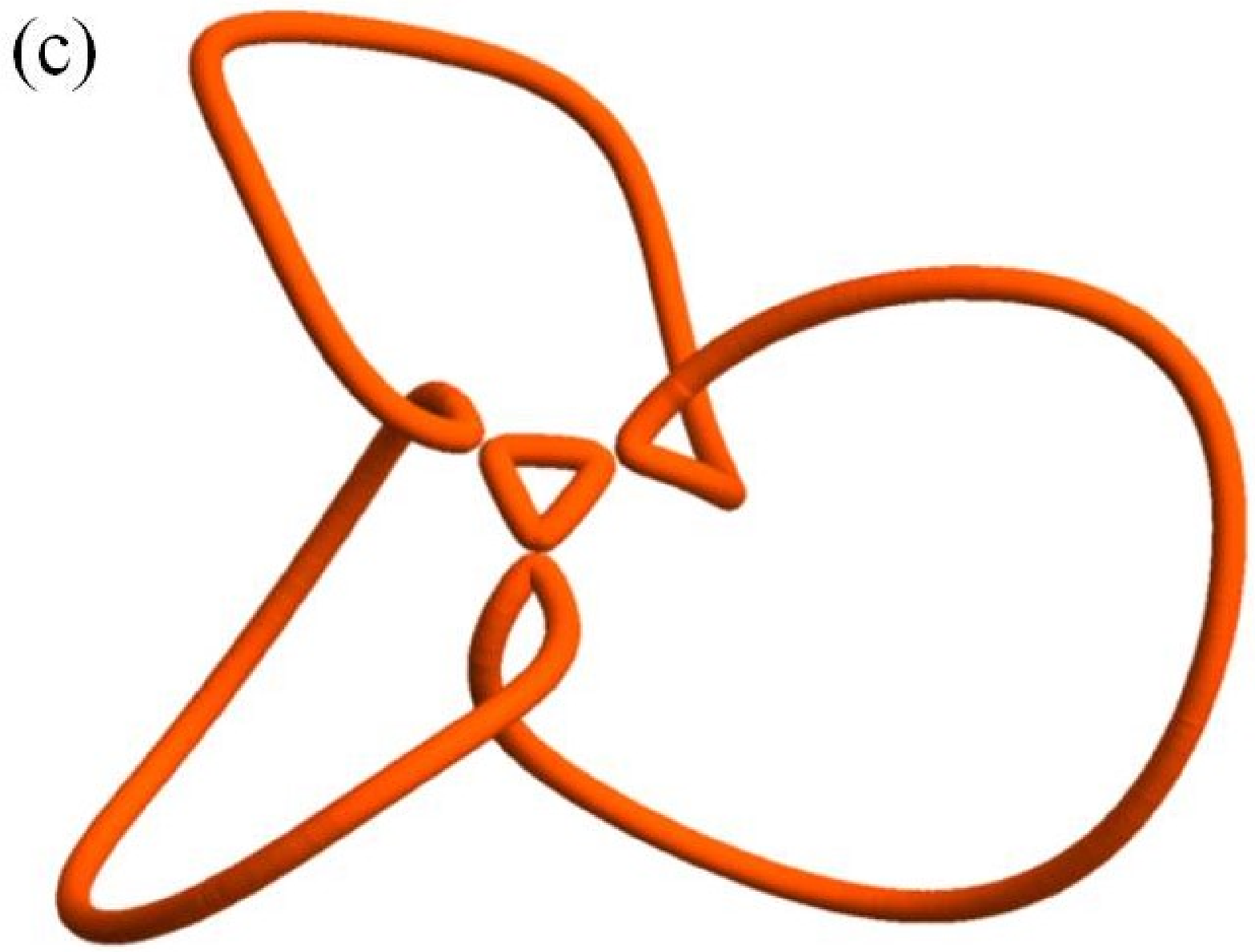}}
\subfigure{\includegraphics[width=4cm, height=3.5cm]{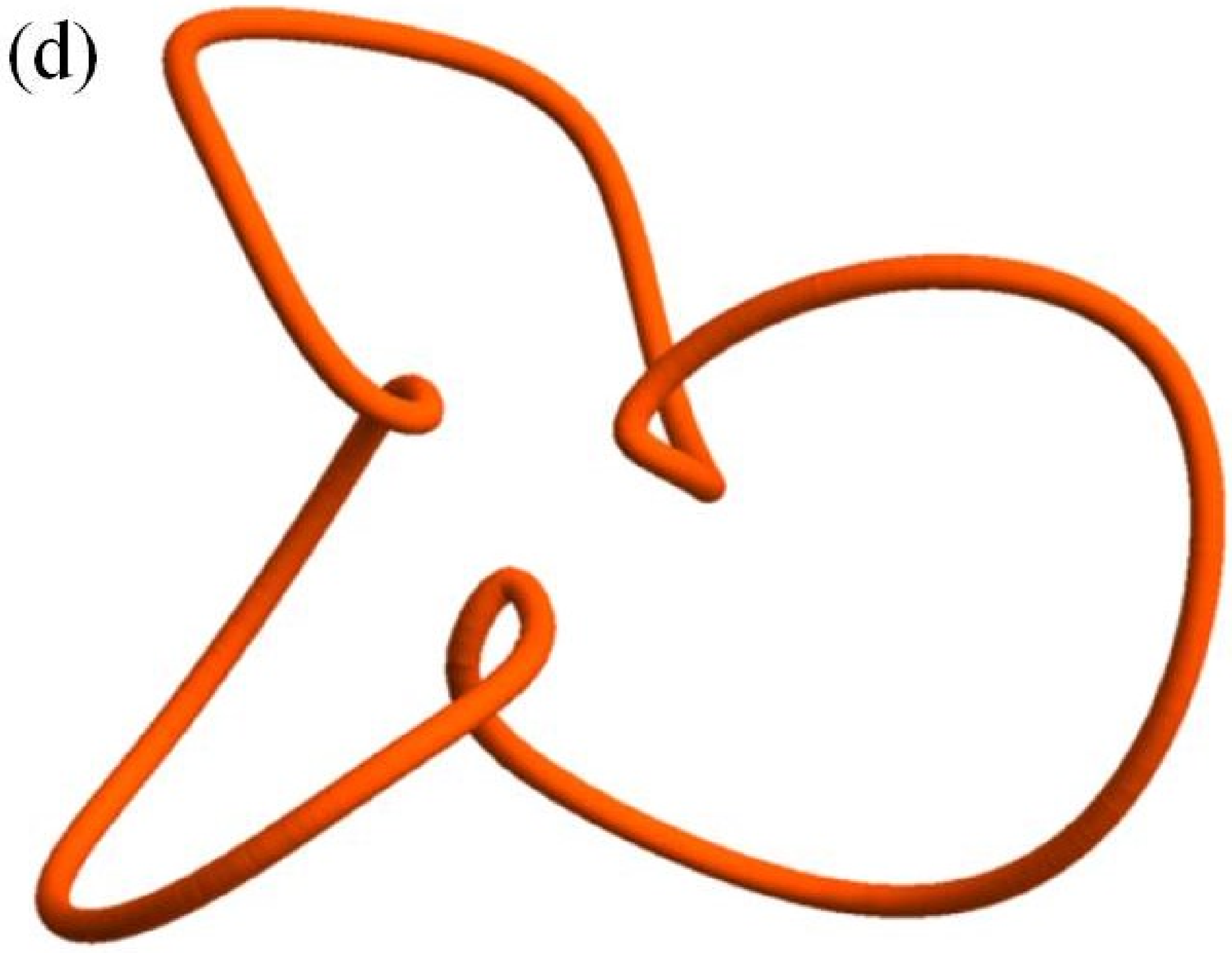}}
\caption{ The evolution of nodal knot as a function of $m_z$. Here, $m_{0}=2.7$ is fixed.   (a) $m_{z}=0.1200$; knotted.
(b) $m_{z}=0.1316$; critical. (c) $m_{z}=0.1335$; unknotted (two rings). (d) $m_{z}=0.1400$; unknotted (single ring).   The knotted-unknotted transition occurs through nodal-line reconnections.  }  \label{transition}
\end{figure}

\begin{figure*}
\subfigure{\includegraphics[width=5.35cm, height=4.8cm]{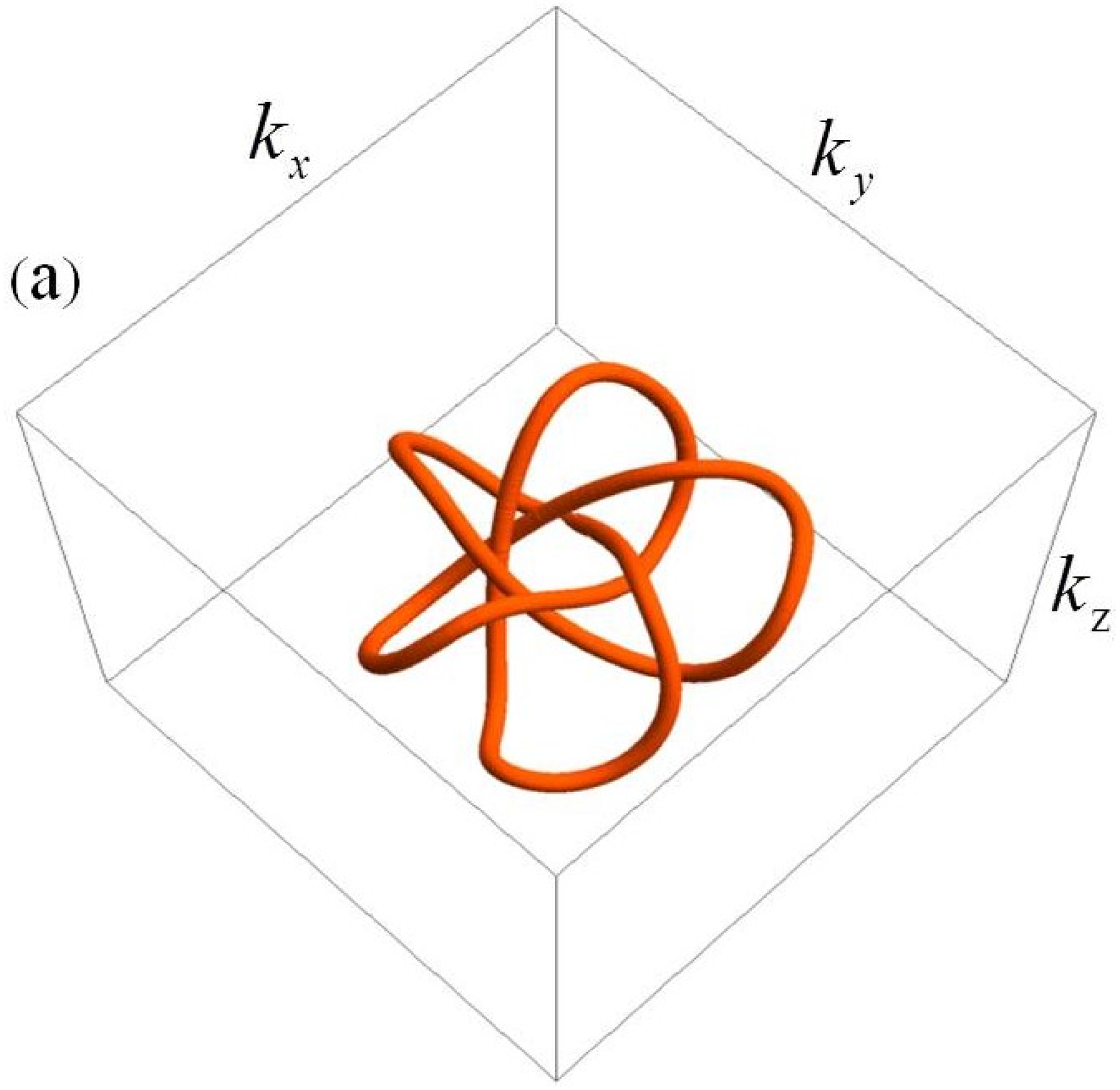}}
\subfigure{\includegraphics[width=5.35cm, height=4.8cm]{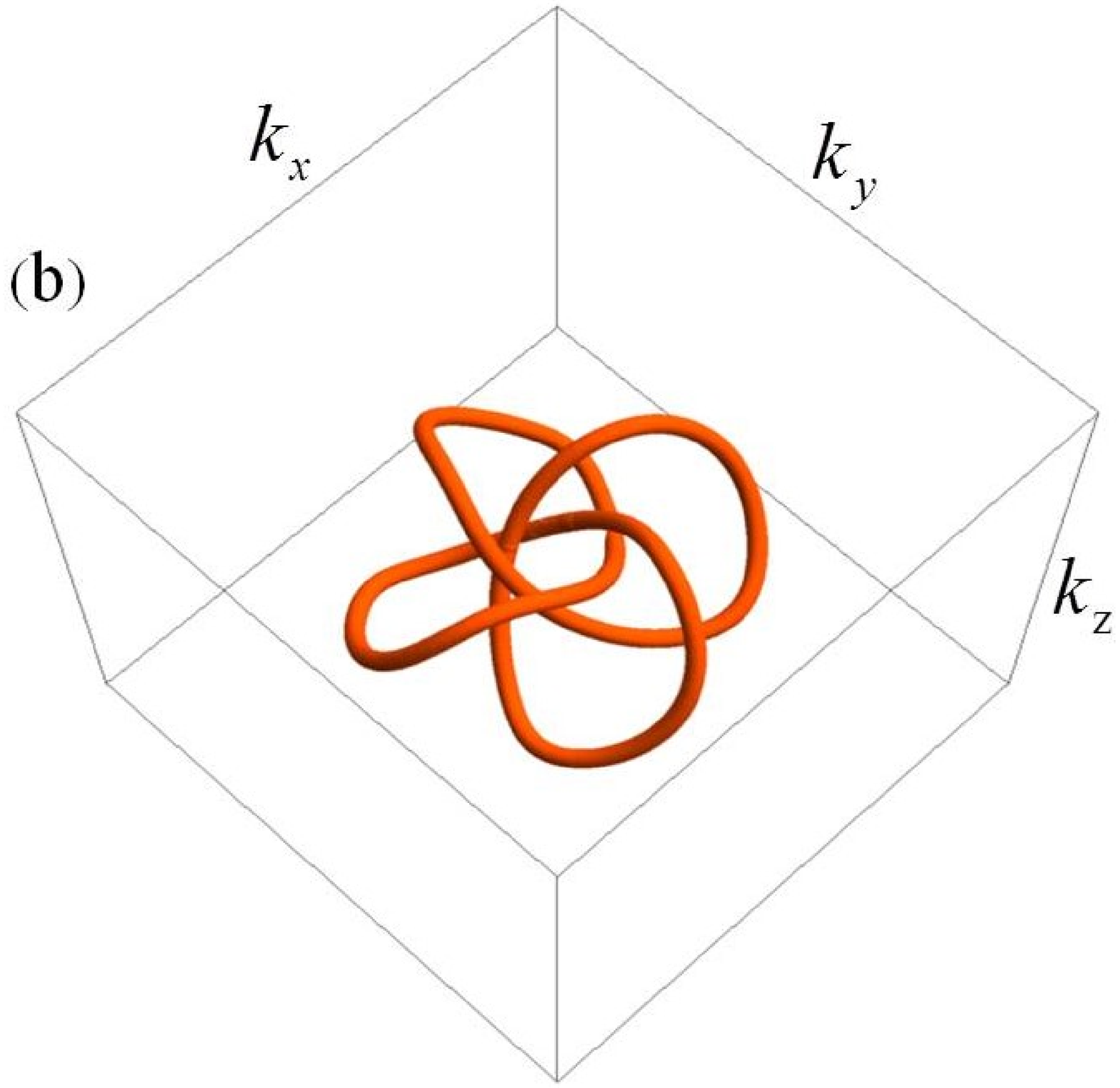}}
\subfigure{\includegraphics[width=5.35cm, height=4.8cm]{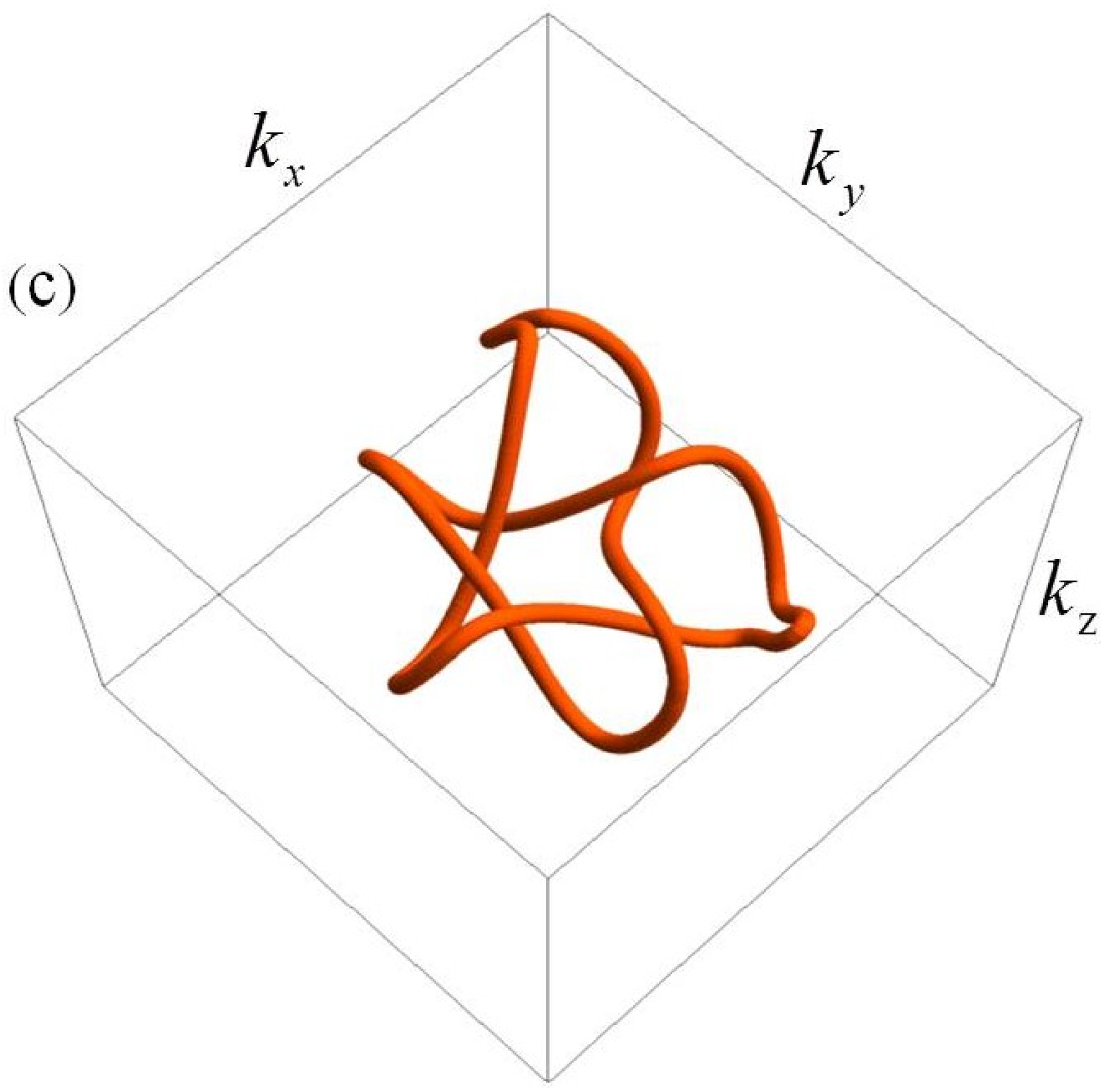}}
\subfigure{\includegraphics[width=5.35cm, height=4.8cm]{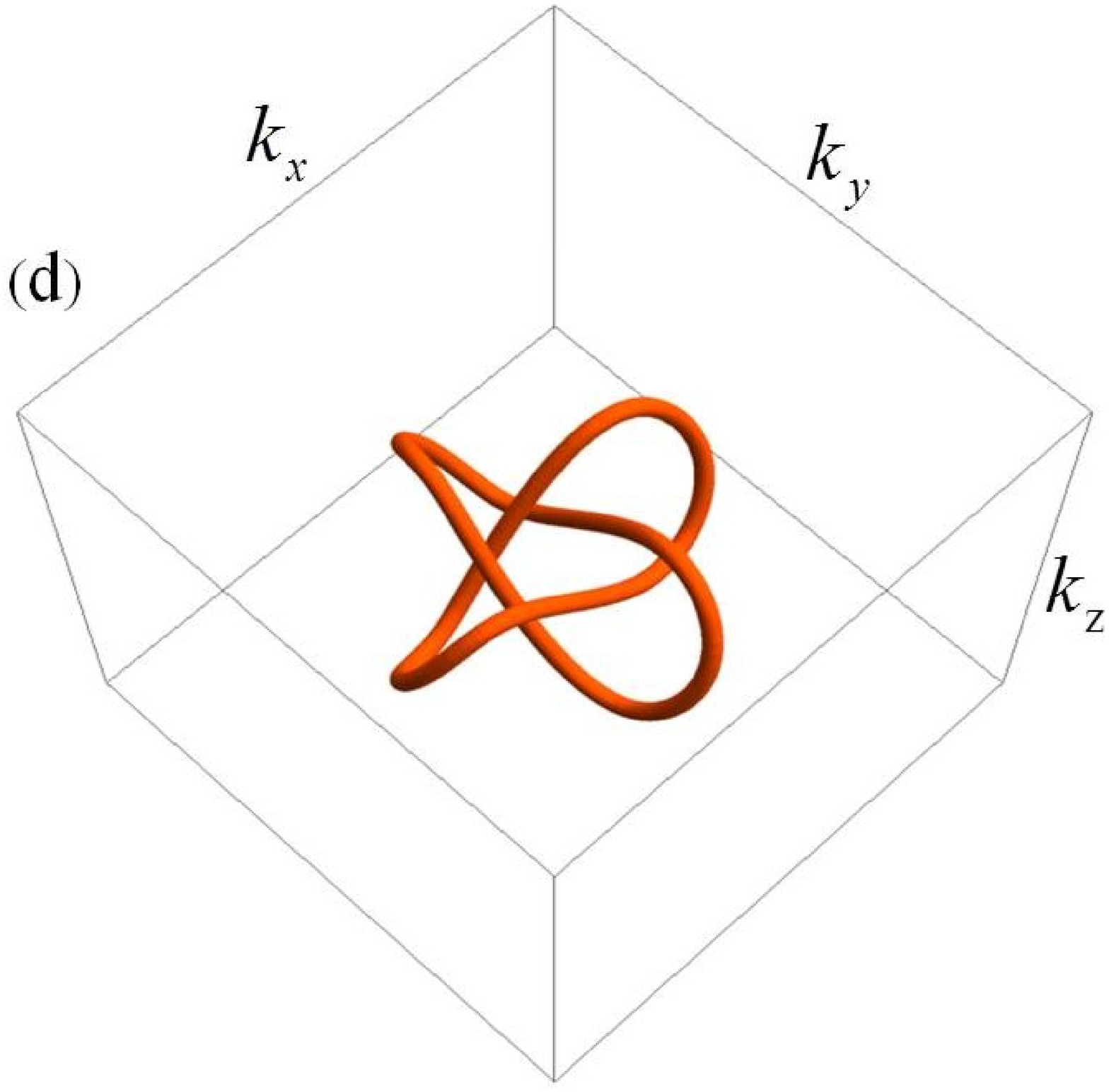}}
\subfigure{\includegraphics[width=5.35cm, height=4.8cm]{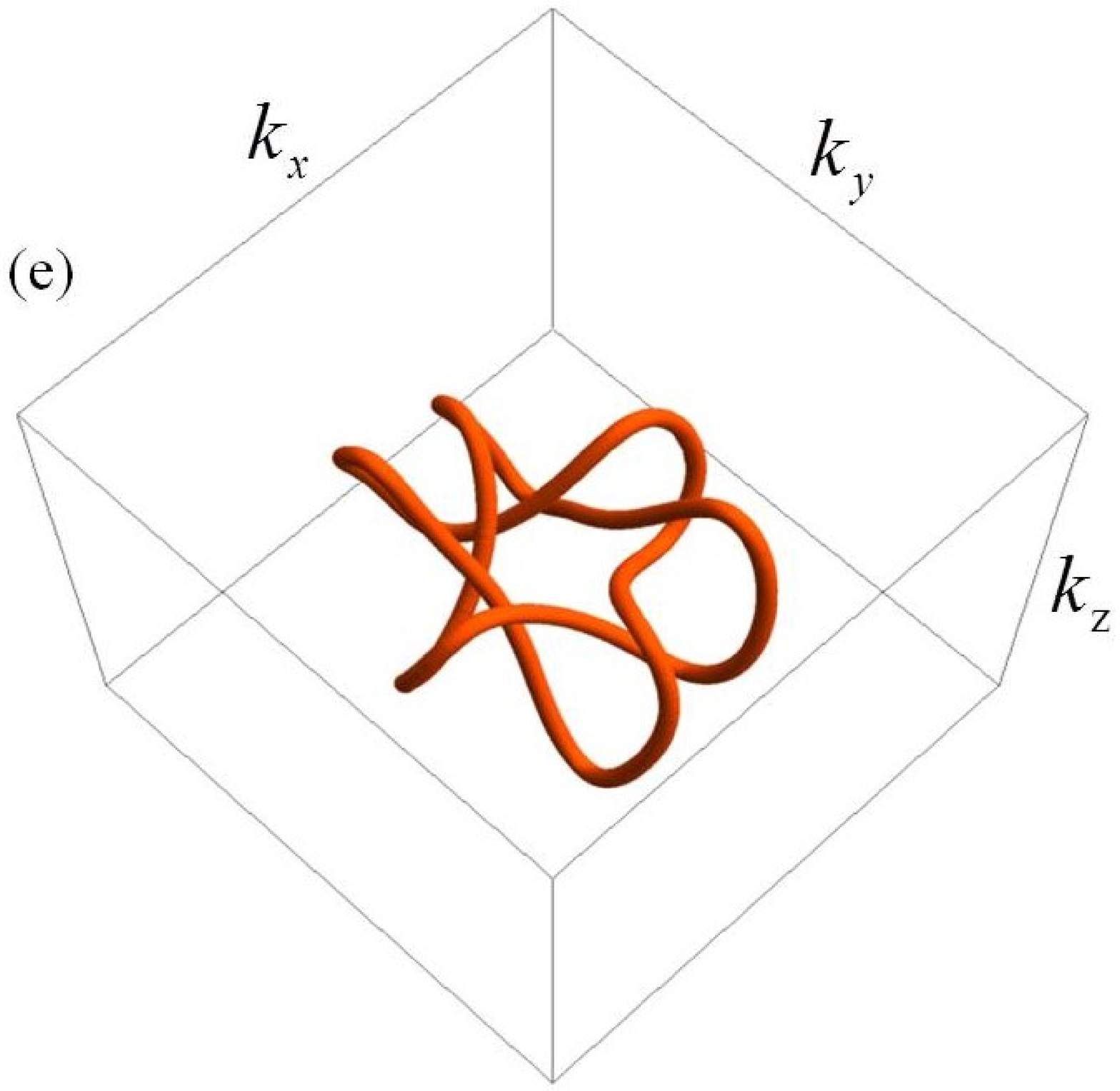}}
\subfigure{\includegraphics[width=5.35cm, height=4.8cm]{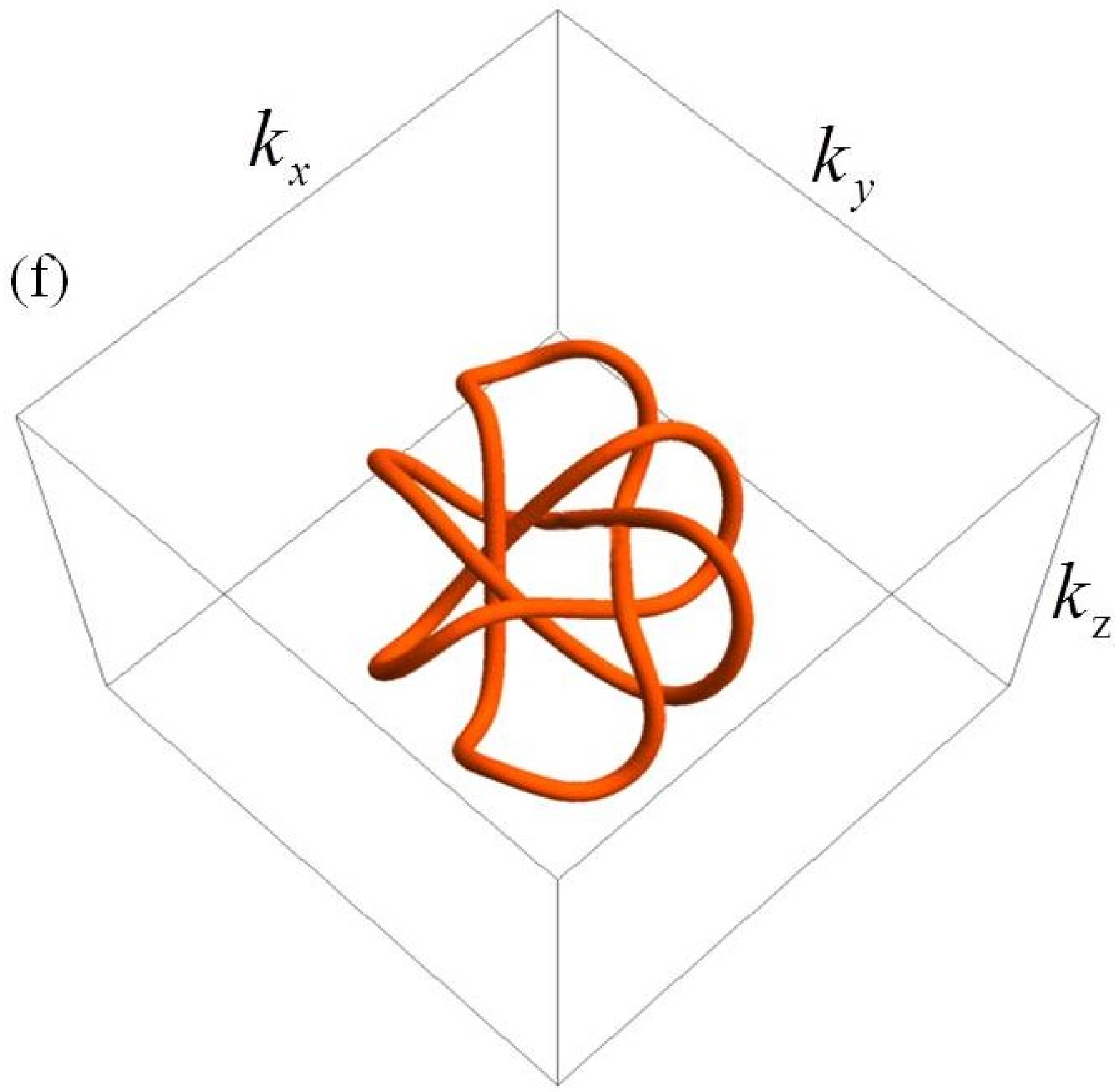}}
\caption{ Nodal knots and nodal links for several values of $(p,q)$. The Hamiltonian is given by the ansatz Eq.(\ref{ansatz}); $m_{0}=2.5$. (a) $(p, q)=(5, 3)$; (b) $(p, q)=(4, 3)$;
(c) $(p, q)=(5, 2)$; (d) $(p, q)=(4, 2)$; (e) $(p, q)=(6, 2)$; (f) $(p, q)=(6, 3)$.  In (a),(b),(c) we have a nodal knot (a single knotted nodal line), while in (d),(e),(f) we have a nodal link. In the cases (a),(b), and (c), $p$ and $q$ are relatively prime; in (d), (e), and (f), they are not. }  \label{other}
\end{figure*}

For lattice models, the $\bk$-space (Brillouin zone) is a 3-torus $T^3$. The main difference compared to the continuum model is that $\bN(\bk)$ must be periodic in $\bk$, otherwise the construction is similar. Following the lead of Eq.(\ref{N}), we can choose
\bea N_1 = \sin k_x, \quad  N_2 = \sin k_y, \quad  N_3 = \sin k_z, \nn  \eea and \bea N_4 = \sum_{i=x,y,z}\cos k_i  -m_0. \label{N2} \eea Expanding $N_4$ in Eq.(\ref{N2}) to the first order, we can see that $3-m_0$ plays the role of $m$. Now Eq.(\ref{ansatz}) gives rise to models of nodal knots in lattice models.  According to Eq.(\ref{ansatz}), the Bloch Hamiltonian for the simplest case $(p,q)=(3,2)$ is given by Eq.(\ref{general}) with \begin{eqnarray}
a_1(\bk)&=&\sin^{2}k_{z}-(\sum_{i}\cos k_{i}-m_{0})^{2}+\sin^{3}k_{x}-3\sin k_{x}\sin^{2}k_{y},\nonumber\\
a_3(\bk)&=&2\sin k_{z}(\sum_{i}\cos k_{i}-m_{0})+3\sin^{2}k_{x}\sin k_{y}-\sin^{3}k_{y}.\quad\quad \label{latticeH}
\end{eqnarray}
It hosts a nodal
knot for $1<m_0<3$ (in this regime, the previously defined winding number $W=-1$). The nodal knot of $m_0=2.5$ is already shown in Fig.\ref{sketch}(d) as a representative of nodal knots. The presence of nodal knots implies the existence of interesting surface flat bands, which are indeed found in our numerical calculations. The regions of surface bands are shown in Fig.\ref{latticesurface}. It is apparent that the $\bk$-space boundary of the surface-state bands is the nodal knot projected to the surface Brillouin zone. As the Hamiltonian has a chiral symmetry (the $\sigma_y$ term is absent), the number of surface zero-energy band is determined by the winding number\cite{Ryu2002} in each region of the surface Brillouin zone.

It is also interesting to investigate the topological transition from the unknotted nodal lines to the knotted ones. To this end, we add an additional $m_z\sigma_z$ term into the Bloch Hamiltonian:  \begin{eqnarray}
a_1(\bk)&=&\sin^{2}k_{z}-(\sum_{i}\cos k_{i}-m_{0})^{2}+\sin^{3}k_{x}-3\sin k_{x}\sin^{2}k_{y},\nonumber\\
a_3(\bk)&=&2\sin k_{z}(\sum_{i}\cos k_{i}-m_{0})+3\sin^{2}k_{x}\sin k_{y} \nn \\ && -\sin^{3}k_{y} + m_z.\quad\quad
\end{eqnarray}   The evolution of nodal knot as a function of $m_z$ is shown in Fig.\ref{transition}.  As we increase $m_z$, the nodal knot gradually deforms. Around $m_z\approx 0.13$,  there are three successive (very close to each other) nodal-line reconnection transitions, taking place at three different locations in the Brillouin zone. At the reconnection transition, two pieces of nodal line come close to each other, touch, and then separate with the lines reconnected. After the reconnections, the original nodal knot evolves to two nodal rings [Fig.\ref{transition}(c)]. As we further increase $m_z$, the smaller ring shrinks and finally disappears. At $m_z=0.14$, we have a single unknotted nodal line [Fig.\ref{transition}(d)].

We have focused on the simplest case $(p,q)=(3,2)$ for simplicity. In fact, the method is general and applicable to all other pairs of integers. As we have explained, nodal knots can be obtained when $(p,q)$ are relatively prime, otherwise nodal links are obtained. We have shown the nodal knots and nodal links for several choices of $(p,q)$ in Fig.\ref{other}. In Fig.\ref{other}(a)(b)(c), different shapes of nodal knots can be found. In contrast, Fig.\ref{other}(d)(e)(f) are nodal links due to the fact that the pairs of integers $(4,2)$ $(6,2)$ and $(6,3)$ are not relatively prime. The even smaller non-relatively-prime integers $(2,2)$ gives a nodal link similar to Fig.\ref{sketch}(c).

\section{Final remarks}

Knots are often studied in the real space. In this paper, we have introduced knots in the momentum space (Brillouin zone), in the context of topological semimetals. Here, the nodal knots emerge as the knotted band-crossing lines.  These topological semimetals have been dubbed ``nodal-knot semimetals'' in this paper, to distinguish them for the ordinary nodal-line semimetals with unknotted nodal lines. We hope that this work can stimulate further applications of the rich subject of knot theory\cite{kauffman2001knots} to topological semimetals.

In this work, we have taken a function of two complex variables to construct nodal knots in three-dimensional Brillouin zone; theoretically, it may be interesting to generalize this method to higher dimensions (with possibly more complex variables).

{\it Note added.--}Upon completing this manuscript, we become aware of a preprint aiming at nodal knots using a different approach\cite{Ezawa2017}. Unlike the everywhere smooth Bloch Hamiltonian here, their trefoil-nodal-knot Hamiltonian has an unavoidable discontinuity in the Brillouin zone (a square-root-type branch cut).

\section{Acknowledgements}

The authors were supported by NSFC under Grant No. 11674189 (R.B., Z.Y., Z.W.),  the National Thousand-Young-Talents Program of China (L.L.), and the Ministry of Science and Technology of China under Grant No.
2016YFA0302400 (L.L.).

\bibliography{dirac}

\end{document}